\documentclass[fleqn,usenatbib]{mnras}
\usepackage{newtxtext,newtxmath}
\usepackage[T1]{fontenc}
\usepackage{ae,aecompl}
\usepackage{graphicx}
\pdfoutput=1
\usepackage{amsmath} 
\usepackage{amssymb}
\usepackage{threeparttable}
\usepackage{float}
\usepackage{epstopdf}

\newcommand{\angstrom}{\textup{\AA}}

\newcommand{\zfourge}{{\sc zfourge}}

\newcommand{\OIII}{[\hbox{{\rm O}\kern 0.1em{\sc iii}}]}
\newcommand{\OII}{[\hbox{{\rm O}\kern 0.1em{\sc ii}}]}

\title[$f_{esc}$ of Extreme \OIII\ Emitters at 
\MakeLowercase{z}$\sim3.5$]{A Low Lyman Continuum Escape Fraction of $<10\%$ for Extreme \OIII\  Emitters in an Overdensity at \MakeLowercase{z}$\sim3.5$}

\author [R.P.~Naidu et al.]{
\parbox[t]{\textwidth}{Rohan P. Naidu$^{1,2}$\thanks{E-mail: rohan.naidu@cfa.harvard.edu}, Ben Forrest$^{3}$, Pascal A. Oesch$^{4}$, Kim-Vy H. Tran$^{3,5}$, and Bradford P. Holden$^{6}$}
\vspace*{8pt}\\
$^{1}$Harvard-Smithsonian Center for Astrophysics, Cambridge, MA 02138, USA.\\
$^{2}$Science Division, Yale-NUS College, 12 College Avenue West, \#01-201, Singapore 138610.\\
$^{3}$George P. and Cynthia W. Mitchell Institute for Fundamental Physics and Astronomy,\\ Department of Physics and Astronomy, Texas A\&M University, College Station, TX 77843, USA.\\
$^{4}$Geneva Observatory, Universit\'e de Gen\``eve, Chemin des Maillettes 51, 1290 Versoix, Switzerland.\\
$^{5}$School of Physics, University of New South Wales, Sydney, NSW 2052, Australia.\\
$^{6}$UCO/Lick Observatory, University of California, Santa Cruz, 1156 High St, Santa Cruz, CA 95064, USA\\
}

\date{Accepted XXX. Received YYY; in original form ZZZ}

\pubyear{2018}

\begin{document}
\label{firstpage}
\pagerange{\pageref{firstpage}--\pageref{lastpage}}
\maketitle

\begin{abstract}
    Recent work has suggested extreme \OIII\ emitting star-forming galaxies are important to reionization. Relatedly, \OIII/\OII\ has been put forward as an indirect estimator of the Lyman Continuum (LyC) escape fraction ($f_{esc}$) at $z\gtrsim4.5$ when the opaque IGM renders LyC photons unobservable. Using deep archival U-band (VLT/VIMOS) imaging of a recently confirmed overdensity at $z\sim3.5$ we calculate tight constraints on $f_{esc}$ for a sample (N=73) dominated by extreme \OIII\ emitters. We find no Lyman Continuum signal ($f_{esc}^{rel} < 6.3^{+0.7}_{-0.7} \%$ at $1\sigma$) in a deep U-band stack of our sample (31.98 mag at 1$\sigma$). This constraint is in agreement with recent studies of star-forming galaxies spanning $z\sim1-4$ that have found very low average $f_{esc}$. Despite the galaxies in our study having an estimated average rest-frame EW(\OIII$\lambda5007$)$\sim400\angstrom$ and \OIII/\OII$\sim 4$ from composite SED-fitting, we find no LyC detection, which brings into question the potential of \OIII/\OII\ as an effective probe of the LyC--a majority of LyC emitters have \OIII/\OII$>3$, but we establish here that \OIII/\OII$>3$ does not guarantee significant LyC leakage for a population. Since even extreme star-forming galaxies are unable to produce the $f_{esc}\sim10-15\%$ required by most theoretical calculations for star-forming galaxies to drive reionization, there must either be a rapid evolution of $f_{esc}$ between $z\sim3.5$ and the Epoch of Reionization, or hitherto observationally unstudied sources (e.g. ultra-faint low-mass galaxies with $\log(M/M_\odot)\sim7-8.5$) must make an outsized contribution to reionization.
\end{abstract}




\section{Introduction}
The protagonists of the last major phase transition of the universe, Cosmic Reionization, have been notoriously elusive. Leaking ionizing radiation into their surroundings, these sources (``Lyman Continuum (LyC) leakers") rapidly turned the universe filled with neutral Hydrogen into one of galaxies in the throes of star-formation and stellar-mass buildup in a mere space of 500 million years \citep{Planck16}. These enigmatic sources are currently expected to be star-forming galaxies, but this has mostly been concluded from elimination of other candidates like quasars and AGN, and hedging on what the faint end (-13 mag) of the UV-Luminosity function may look like rather than through convincing, direct observations of LyC emission from star-forming galaxies (e.g. \citet{Micheva16, Cristiani16, Robertson13, Robertson15, Bouwens16b}, but see \citet{Giallongo15, madau15}). Since ionizing photons cannot directly be observed during the thick of reionization at $z>6$ due to the opaque intervening inter-galactic medium (IGM), observational attempts to understand reionization have resorted to searching for analogues of LyC leakers at lower redshifts \citep{Inoue14, Stark16}.

The prevailing paradigm from various reionization calculations is that for star-forming galaxies to drive reionization, their $f_{esc}$ must exceed $10-15\%$ \citep[e.g.,][]{Mitra15,Mitra16,Giallongo15,madau15,Price16,Feng16}. Initial estimates of the average $f_{esc}$ of star-forming galaxies at $z\sim3-4$ published numbers exceeding this threshold, but it is now well-established that these early studies greatly underestimated how contaminated they were by low$-z$ interlopers and near-neighbor sources \citep{Vanzella10, Vanzella12, Nestor11, Siana15, Mostardi15, Grazian16}. The current, well-established picture is that of extremely humble escape fractions out to $z\sim1-4$ \citep[e.g][]{Rutkowski16, Rutkowski17,Japelj17,Matthee16,Marchi16, Grazian16, Grazian17}.

Meanwhile, major strides have been made in compiling a sample of individual LyC leakers. The COS instrument on $HST$ has been effective at finding these sources in our local universe and out to $z\sim0.3$ \citep{Bergvall06,Leitet13, Izotov16a, Izotov16b, Borthakur14, Leitherer16}. And for the first time, we have a sample of confirmed LyC leakers and highly likely candidates at the higher redshifts of $z\sim2-3.2$ \citep{Shapley16, Vanzella16, Bian17,Mostardi15,Naidu17}. Using this sample, we hope to understand the mechanisms of LyC escape, and robustly link $f_{esc}$ to quantities that are observable in the Epoch of Reionization.

One such indirect measure of $f_{esc}$ that has been put forward in the literature is the line ratio \OIII/\OII\ which potentially traces density-bounded HII regions \citep[e.g.,][]{ Nakajima16, Faisst16}. A large fraction of confirmed LyC leakers and highly likely candidates show extreme \OIII/\OII\ ($\gtrsim3$)  (five sources at $z\sim0.3$ from \citet{Izotov16a,Izotov16b}, $Ion2$ at $z\sim3.2$ from \citet{deBarros15}, three sources at $z\sim2$ from \citet{Naidu17}). Many of these sources also display extreme EW(\OIII), in some cases as high as $\sim1000\angstrom$ (rest-frame) (e.g. $Ion2$, GS-30668 from \citet{Naidu17}). Extreme \OIII\ emission has long been suspected to imply LyC emission, as in the case of the compact, low-metallicity ``green pea" galaxies in the SDSS sample which have been studied extensively for this reason \citep{Cardamone09, Jaskot13,Nakajima14, Henry15, Amorin17}.

However, the $f_{esc}$-\OIII/\OII\ connection and the contribution of extreme \OIII\ emitters to the average LyC emission is yet to be tested observationally on statistical samples. In this work, we leverage a recently confirmed overdensity at $z\sim 3.5$ \citep{Forrest17} to assemble a sample in a narrow redshift bin that is dominated by high EW(\OIII) sources. We use this unique sample to constrain $f_{esc}$ at $z\sim3.5$, and also test the viability of \OIII/\OII\ and EW(\OIII) as indirect handles on $f_{esc}$. This paper is structured as follows. In section \S2 we describe the imaging and redshift data used for the analysis. The methodology of our stacking and calculation of $f_{esc}$ is outlined in \S3. In \S4 we present constraints on $f_{esc}$ for various substacks, and in \S5 we discuss the implications of our findings on the \OIII/\OII-LyC connection in the context of other reionization studies. Finally, in \S6 we summarise our work and look to the future. Throughout this paper, we adopt $\Omega_M=0.3, \Omega_\Lambda=0.7, H_0=70$ kms$^{-1}$Mpc$^{-1}$, i.e., $h=0.7$, largely consistent with the most recent measurements from Planck \citep{Planck2015}. Magnitudes are given in the AB system \citep{Oke83}. All limits on the escape fraction mentioned in this paper are $1 \sigma$ limits, unless stated otherwise. \OIII/\OII\ as used in this paper denotes the \OIII$\lambda$5007\angstrom/\OII$\lambda\lambda 3727,3729\angstrom$ line ratio. We use $f_{esc}$ to denote the absolute escape fraction, and $f_{esc}^{rel}$ denotes the ``relative" escape fraction (defined in \S3.2). All Equivalent Widths (EWs) mentioned in this paper are in the rest-frame.

\begin{figure}
\centering
\includegraphics[width=0.95\linewidth]{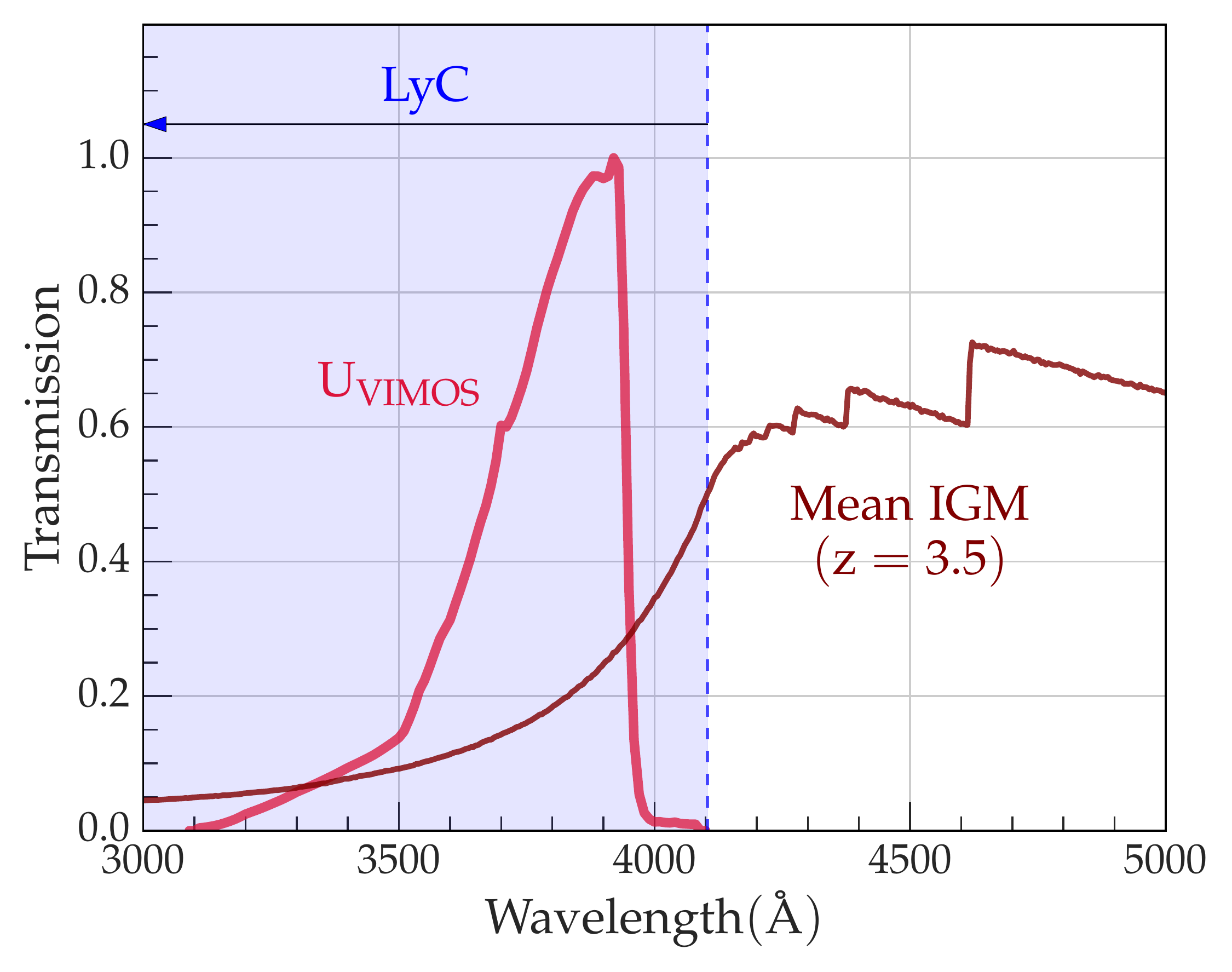}
\caption{At $z=3.5$ the VLT/VIMOS U-band (transmission curve shown in red) probes exclusively LyC photons ($\lambda$(restframe)$<912\angstrom$, shown in blue) which makes it ideal for a study constraining $f_{esc}$. Plotted in magenta is the mean IGM transmission from \citet{Inoue14} obtained by averaging 10,000 lines of sight at $z=3.5$.}
\label{fig:filter+igm}
\end{figure}

\section{Data}
\label{sec:data}
\subsection{Photometry}
\subsubsection{Deep U-band and R-band VLT/VIMOS imaging}

In this study we use deep U-band and R-band imaging from \citet{Nonino09} (27.9 mag and 27.5 mag deep respectively, at $5\sigma$ in a $1.2''$ aperture). This imaging dataset has already been used to make successful forays into the LyC puzzle. The discovery of $Ion2$, the first definitively confirmed LyC leaker at $z>3$ \citep{Vanzella15}, and one of the preliminary constraints on $f_{esc}$ at $z\sim 3-4$ \citep{Vanzella10b} were possible due to the depth of these images. In our case, at $z\sim3.5$, the U-band contains exclusively Lyman Continuum photons while the R-Band contains non-ionizing UV flux (see Figure \ref{fig:filter+igm}).

\subsubsection{HST imaging}
Several previous studies of the LyC using ground-based data suffered from contamination due to foreground sources that was often reported as an offset between the non-ionizing UV flux and the purported LyC flux (see \citet{Vanzella10} for a definitive account). In order to sidestep this pitfall we rely on high spatial resolution Hubble Space Telescope images of the galaxies in our sample to check for contamination from nearby neighbours and low$-z$ interlopers. Using $HST$ images to discount contamination has become a standard step in recent LyC studies \citep[e.g.][]{Matthee16, Japelj17, Shapley16}.

For our galaxies, $HST$ images are sourced from the 3D$-$HST \citep{Skelton14, Brammer12, Momcheva16} and CANDELS \citep{candels, candels2} surveys, as well as the previous GOODS ACS imaging \citep[][]{Giavalisco04a}. 60 out of the 73 galaxies that comprise the sample studied in this paper have been imaged in the F435W, F606W, F775W, F814W, F125W, F140W, and F160W $HST$ filters by the aforementioned surveys, and the remaining 13 have been observed using at least five of these filters.

\begin{figure}
\centering
\includegraphics[width=0.95\linewidth]{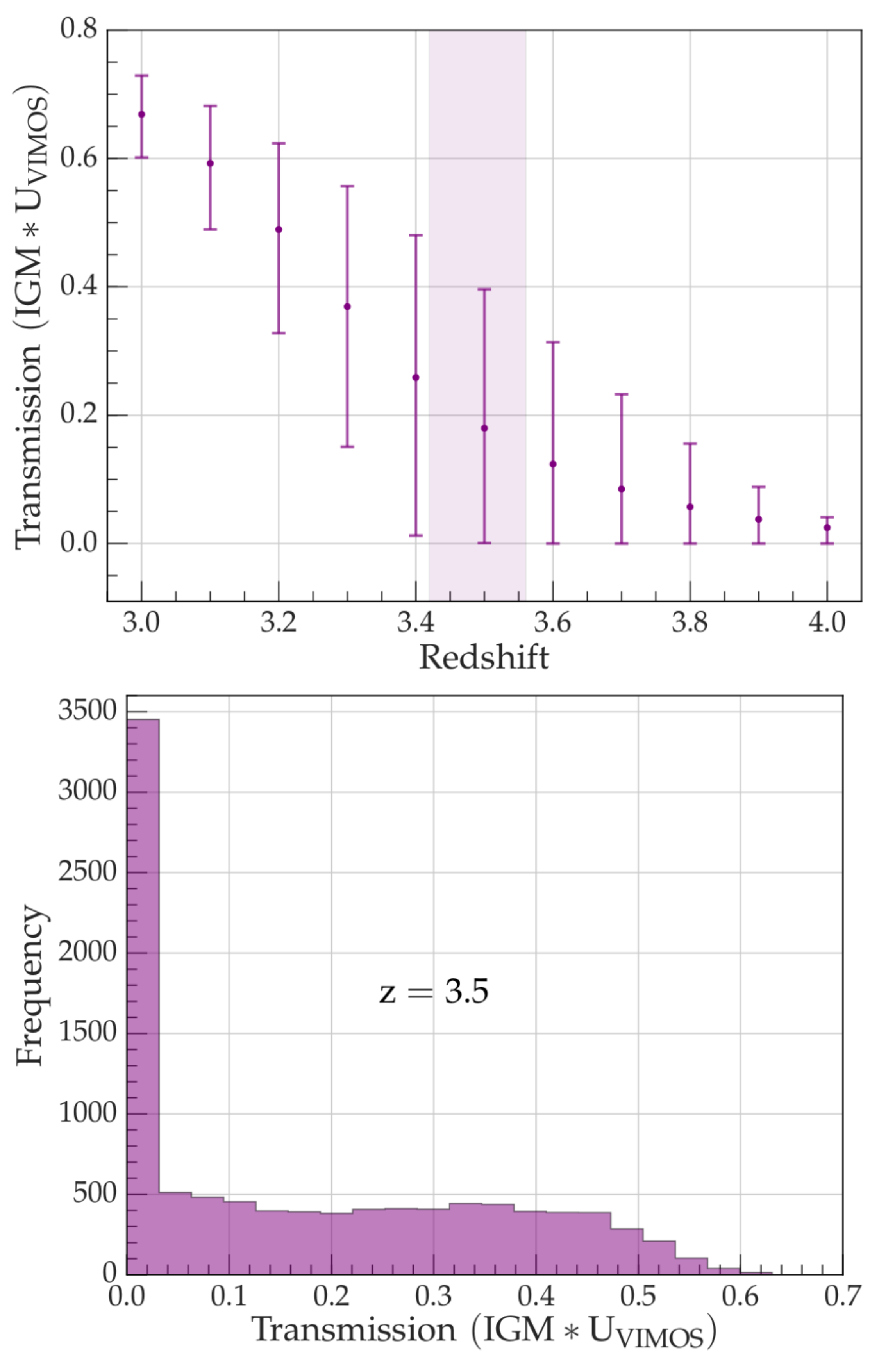}
\caption{\textbf{Top:} Mean IGM transmission convolved with the U-band plotted against redshift. Error-bars indicate $68\%$ ranges of the distribution, and the shaded region represents the redshift range of our sample. Stacking galaxies across a wide redshift-range (e.g. $z=3-4$) to calculate $f_{esc}$, as done in several previous studies, introduces uncertainties due to the rapid evolution of the IGM with redshift. Since our entire sample lies in a narrow slice ($z=3.42-3.58$) we are able to avoid this. \textbf{Bottom:} The transmission of 10,000 IGM lines of sight from \citet{Inoue14} convolved with the U-band at $z=3.5$. The mean transmission is $0.18_{-0.18}^{+0.22}$, and for $\sim 3800$ lines of sight the transmission is $<0.05$, while for $\sim100$ the transmission is zero. The large variance in transmission reflects the stochastic nature of the IGM, and points to the need for large samples over which the variance along individual lines of sight may be averaged out.}
\label{fig:igm_zevolution}
\end{figure}

\subsection{Redshifts} 
\subsubsection{Overdensity at $z\sim3.5$}
\citet{Forrest17} presented an overdensity in the CDFS field based on photometric redshifts derived from the medium band near-infrared FourStar Galaxy Evolution Survey \citep[\zfourge;][]{Straatman16}. The overdensity is detected at $10 \sigma $ significance and existing literature spectroscopic redshifts validate the detection. The photometric redshifts derived by \zfourge\  using EAZY \citep{Brammer08} are made particularly reliable by the survey's use of the rich multiwavelength legacy data available in the CDFS field (40 filters spanning the UV to mid-IR). Further, the medium band filters of the \zfourge\  survey finely sample the NIR which contains the distinctive $4000\angstrom$ Balmer break at $z>1.5$, thus preventing several low-z interlopers making it into our $z\sim 3.5$ selection. The sources from the overdensity featured in this paper have photometric redshifts derived from 38 bands on average, with a mean uncertainty of $\sim 0.01$ ($\sigma_{z}/(z+1)$).

By fitting composite SEDs and inferring nebular line fluxes from K-band excesses for individual galaxies, \citet{Forrest17} show that the overdensity includes galaxies that are ``Strong Emission Line Galaxies (SELGs)" and ``Extreme Emission Line Galaxies (EELGs)". The EELGs and SELGs in \citet{Forrest17} have on average an EW(\OIII+H$\beta$) of  $\sim 803 \pm 228\angstrom$ and $\sim 230 \pm 90 \angstrom$ respectively. The sample studied in this paper is comprised largely of sources that belong to these two classes ($74\%$ of the selected sample). 

\subsubsection{Spectroscopic Redshifts}
We also select galaxies from public redshift catalogs that have spectroscopic redshifts in the redshift-range of the overdensity ($z=3.42-3.58$), and that also have coverage in the deep U-Band and R-Band imaging from \citet[][]{Nonino09}. The redshift catalogs we use are the compilation in the 3D-HST photometric release \citep{Skelton14}, the VUDS survey \citep[DR1,][]{vuds} and the ESO GOODS/CDFS master redshift catalogue\footnote{\url{https://www.eso.org/sci/activities/garching/projects/goods/MasterSpectroscopy.html}}, compiled mainly from \citet{Vanzella08}, \citet{Lefevre05}, and \citet{Balestra10}. We only consider spectroscopic redshifts which have the highest quality flags (e.g confidence class 3+ from the VUDS survey). We also use MOSFIRE redshifts from the sample studied in \citet{Holden16}.

\begin{figure}
\centering
\includegraphics[width=0.95\linewidth]{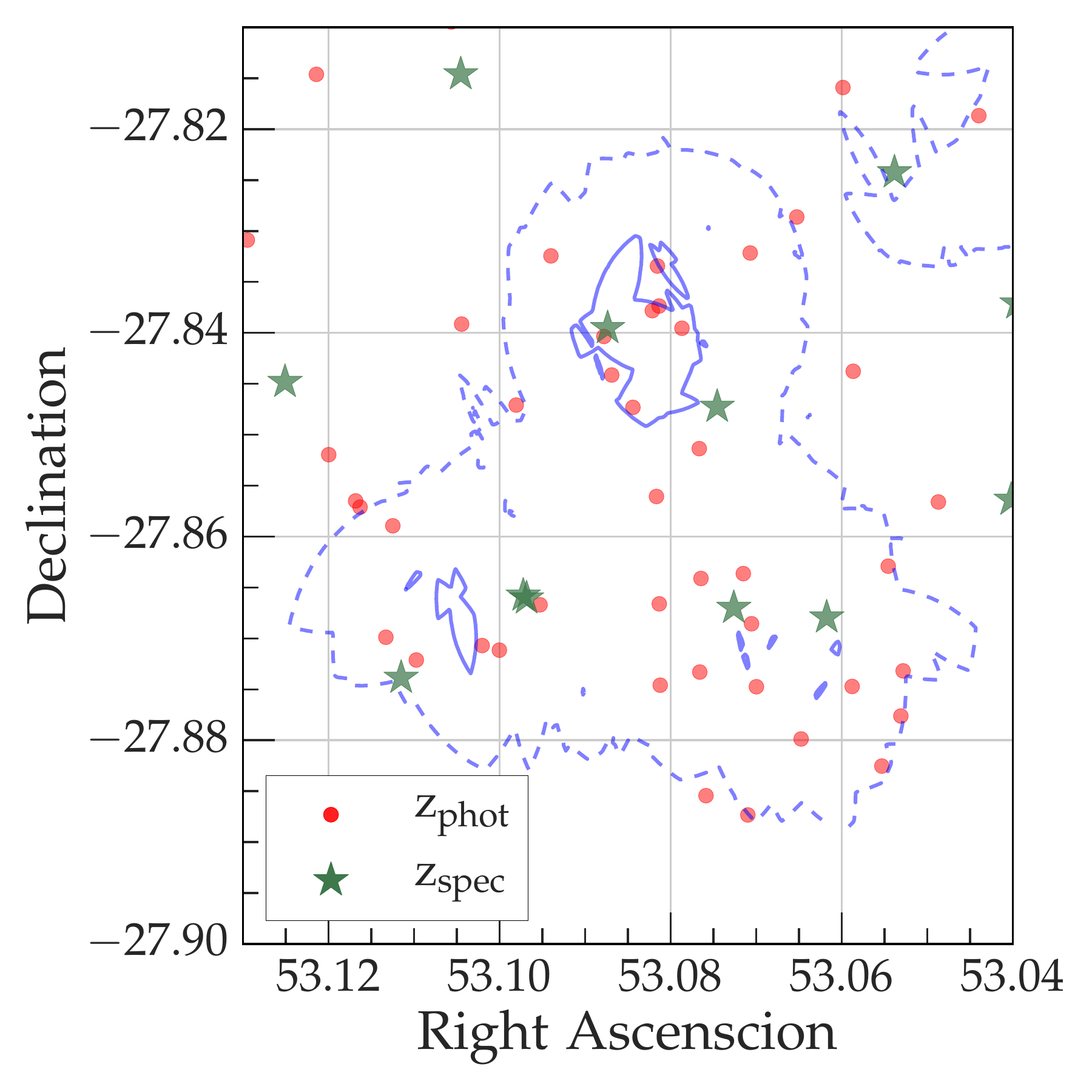}
\caption{Spatial distribution of \zfourge\ galaxies at $3.42 \leq z \leq 3.56$ included in our sample. The contours in blue, as computed in \citet{Forrest17}, represent $1\sigma$ (dashed lines) and $5\sigma$ (solid lines) deviations from the mean density of the field. This clustering of sources in a narrow $z$-range helps us assemble a sample uniquely suited to studying the LyC, since we are able to avoid uncertainties associated with the rapid $z$-evolution of the IGM (see Figure \ref{fig:igm_zevolution})
}
\label{fig:spatial_dist}
\end{figure}

\section{Methodology}
\label{sec:methods}

In this section we discuss how we compiled our sample from the data discussed in the previous section, our procedure for stacking the deep U and R band images of the selected sample, and finally how we calculate the average $f_{esc}^{rel}$ of our stack.

\begin{figure}
\centering
\includegraphics[width=0.95\linewidth]{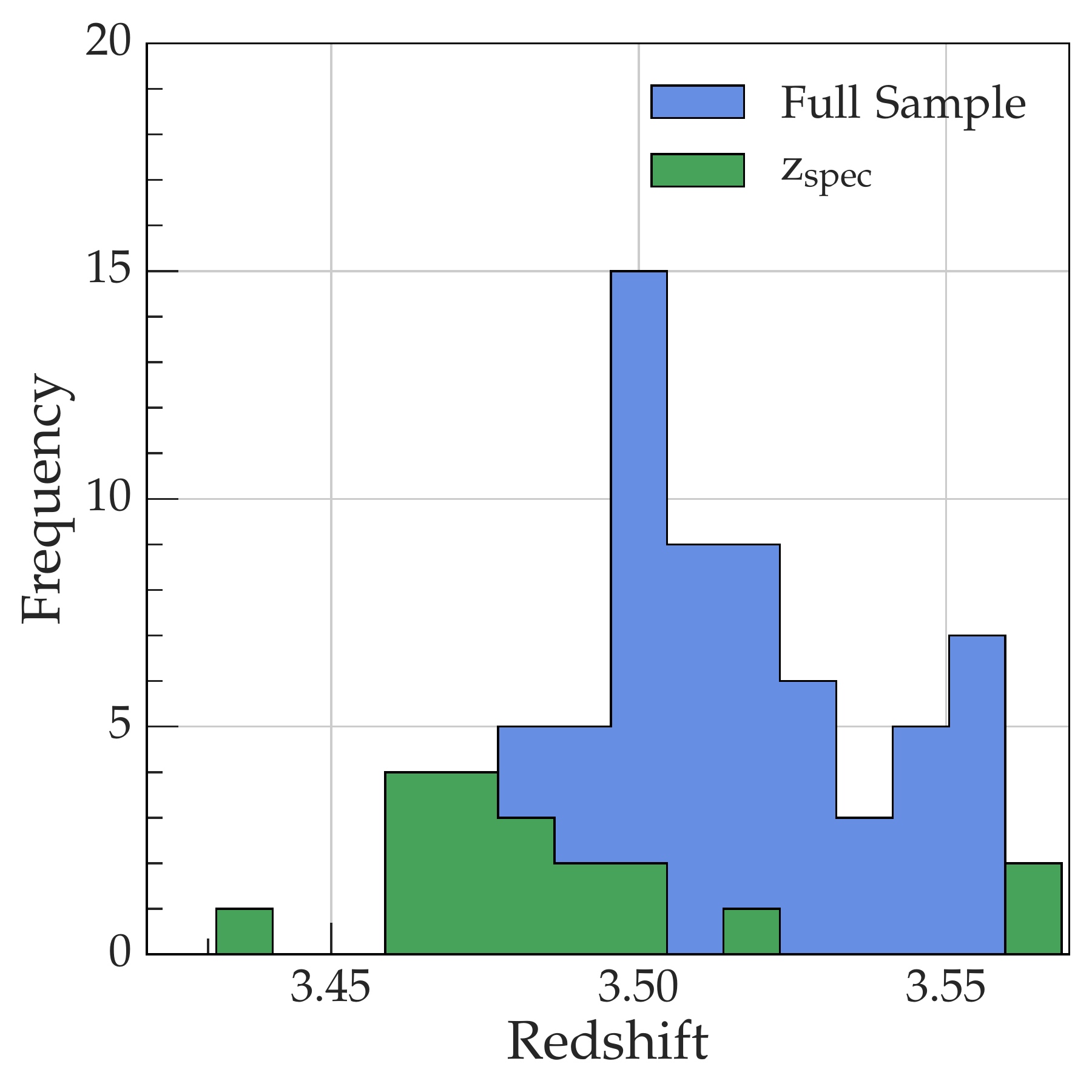}
\caption{Redshift distribution of the selected sample (N=73) studied in this paper. 54 sources have $\geq32$-band photometric redshifts and 19 have spectroscopic redshifts. A sizeable sample like this one is required to counter the large variance of the IGM by averaging over a large number of sight lines.
}
\label{fig:redshift_dist}
\end{figure}

\subsection{Sample Selection}
Stacking LyC-containing images of galaxies across a wide redshift-range to calculate $f_{esc}$ introduces uncertainties due to the rapid evolution of the IGM with redshift (shown in Figure 2). Further, due to the stochasticity of the IGM, a large sample is required to average out the biases along individual lines of sight. Of all the efforts to constrain $f_{esc}$ at $z>3$ by stacking photometry, to our knowledge only \citet[][]{Grazian16,Grazian17} launched a targeted campaign to compile a sample that is located in a narrow enough redshift-bin ($z=3.27-3.40$) to obviate IGM evolution with redshift, is of a large enough size (69 galaxies) to account for the stochastic variation of the IGM, and is free of contamination with a high degree of confidence based on HST coverage.

Thanks to the overdensity discussed earlier, we are able to put together a similarly large sample (73 galaxies) in a very narrow redshift-bin at the slightly higher redshift of $z=3.42-3.58$.  We choose this range since it contains the majority of the overdensity galaxies, and is a range at which the U-band is completely covered by the LyC. In order to make it to our sample, we require the peak photometric redshift of an overdensity galaxy to lie in $z=3.42-3.58$ and that the $5 \sigma$ lower limit on the redshift be greater than $z=3.35$ since at $z < 3.35$, non-ionizing UV radiation begins leaking into the U-band. In addition to the overdensity galaxies, we also consider galaxies in the \citet{Nonino09} imaging with literature spectroscopic redshifts at $z=3.42-3.58$ for inclusion in our sample. 

We find no evidence for any source in this sample being an AGN. For this we rely on the \zfourge\  AGN catalogue \citep[][]{Cowley16}. In addition, we also use latest x-ray \citep[][]{Xue16, Cappelluti16} and variability-based AGN catalogues \citep{Villforth10, Sarajedini11, Garcia-Gonzalez15}.

Next, like in \citet[][]{Japelj17} and \citet[][]{Marchi16} we make VJH RGB stamps (V-J, J-H, and V-H comprising the three channels of the RGB images) using the high spatial resolution F606W, F125W, and F160W images and inspect these stamps to exclude galaxies that show signs of photometric contamination from nearby neighbours, or of blending with low$-z$ interlopers. In the VJH stamps, flux contamination in all except some pathological cases shows up with a colour different from that of the central source.

After making these cuts, we are left with a sample of 73 galaxies, 19 of which have secure spectroscopic redshifts from the literature, while the remaining 54 have $\geq32$-band \zfourge\  photometric redshifts. The sample is also dominated by SELGs and EELGs as classified by \citet{Forrest17} (54 out of 73)--this aspect is discussed further in \S\ref{sec:Discussion}.

\begin{figure}
\centering
\includegraphics[width=0.75\linewidth]{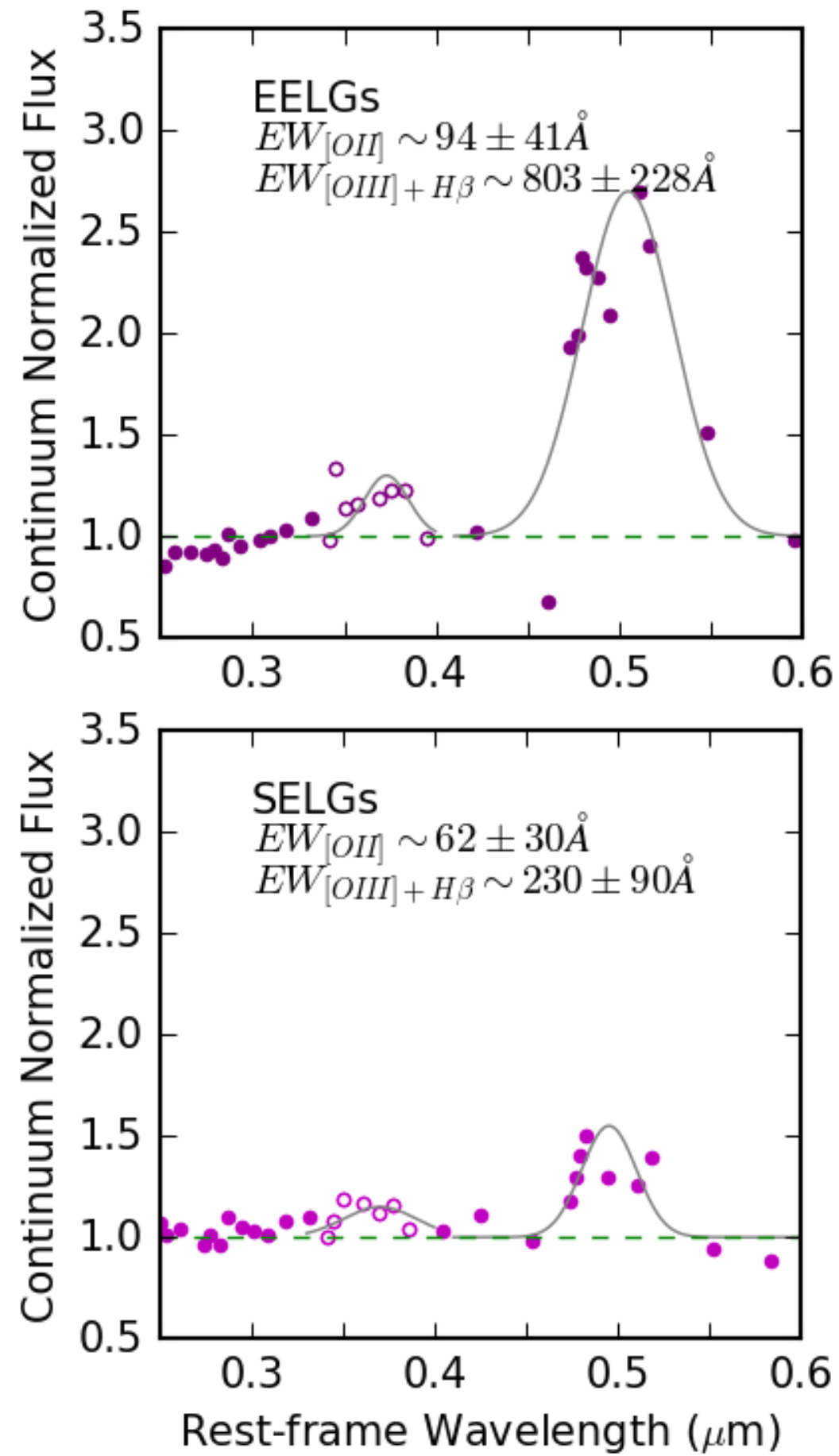}
\caption{Composite SEDs for galaxies classified as Extreme Emission Line Galaxies (EELGs) and Strong Emission Line Galaxies (SELGs) from \citet{Forrest17}. Each point represents the median of a number of de-redshifted and scaled photometric measurements in a narrow wavelength bin. Empty circles are those associated with \OII, while filled circles determine the continuum and \OIII. 54 of the 73 sources in our study belong to these two classes, and we refer to them as the ``EELG" sub-sample. \OIII$\lambda\lambda4959,5007$+H$\beta$ and \OII\ are visible in these fits, and we use their estimated EWs to get a sense of \OIII/\OII\ for our sample.}
\label{fig:oxygen_ratio}
\end{figure}

\subsection{Calculating $f_{esc}$}

We use the following equation for $f_{esc}^{rel}$ \citep[][]{Steidel01, Siana07}: 

\begin{equation}
f_{esc}^{rel} = \frac{f_{U}/f_{R}}{L_{900}/L_{1500}}\frac{1}{<T(IGM)_{U}>}
\end{equation}

Most studies of LyC photon escape measure the relative escape fraction as defined above instead of the absolute escape fraction to avoid uncertainties that arise from estimating the dust attenuation ($f_{esc} = f_{esc}^{rel} 10^{-0.4 A_{UV}} $, $f_{esc} \leq f_{esc}^{rel}$). In the rest of this section we describe how we treat each term in Equation 1.

$f_{U}/f_{R}$ is the observed flux ratio of the Lyman Continuum flux and non-ionizing UV flux that we observe in the U-band and R-band stacks respectively. At $z=3.5$ the U-band collects flux at restframe $\lambda\leq900\angstrom$ and the R-band covers restframe $\lambda=1280-1580\angstrom$.

$L_{900}/L_{1500}$ is the intrinsic luminosity density ratio. No definitive observational constraints exist for this quantity, and it is usually derived by averaging SED models. We set this ratio to $0.2$ motivated by {\tt Starburst99} models \citep{Leitherer99} consistent with our EELG sample, for which we assume \OIII$\lambda5007$/H$\beta\sim5.1$ and fit EW(\OIII$\lambda5007)\sim400\angstrom$. These values correspond to EW($H\beta$)$\sim60\angstrom$ and thus to an age of $\sim5$ Myr for the instantaneous burst, and to $\sim100$ Myr for continuous star formation with a constant star formation rate. The models yield $\mathrm{L_{900}/L_{1500}} \sim 0.15-0.20$. Several previous studies assume comparable values \citep[e.g. $\sim1/3$ in][]{Marchi16, Grazian16, Guaita16, Grazian17, Japelj17}. In our own previous work \citep{Naidu17} we have deployed a grid of BPASSv2 models \citep{Eldridge17} which span $\mathrm{L_{900}/L_{1500}} \sim 0.1-0.5$. We also note that the $\mathrm{L_{900}/L_{1500}}$ ratio has been observed to be higher than what we assume ($\sim0.36$) for some more extreme, individual low-$z$ LyC leakers \citep[e.g.][]{Izotov16b}.

$<T(IGM)_{U}>$ represents the mean transmission of the IGM at $z=3.5$. We calculate $<T(IGM)_{U}> =  0.18_{-0.18}^{+0.22}$ by convolving the U-band's filter transmission curve with 10,000 IGM lines of sight at $z=3.5$ from \citet{Inoue14}. The top panel of Figure \ref{fig:igm_zevolution} shows the rapid evolution of $<T(IGM)_{U}>$ with redshift, highlighting the importance of the narrow redshift distribution of our sample. The bottom panel of Figure \ref{fig:igm_zevolution} shows the stochastic nature of the IGM and the large variance in transmission across the 10,000 lines of sight at $z=3.5$.

\begin{figure}
\centering
\includegraphics[width=1.05\linewidth]{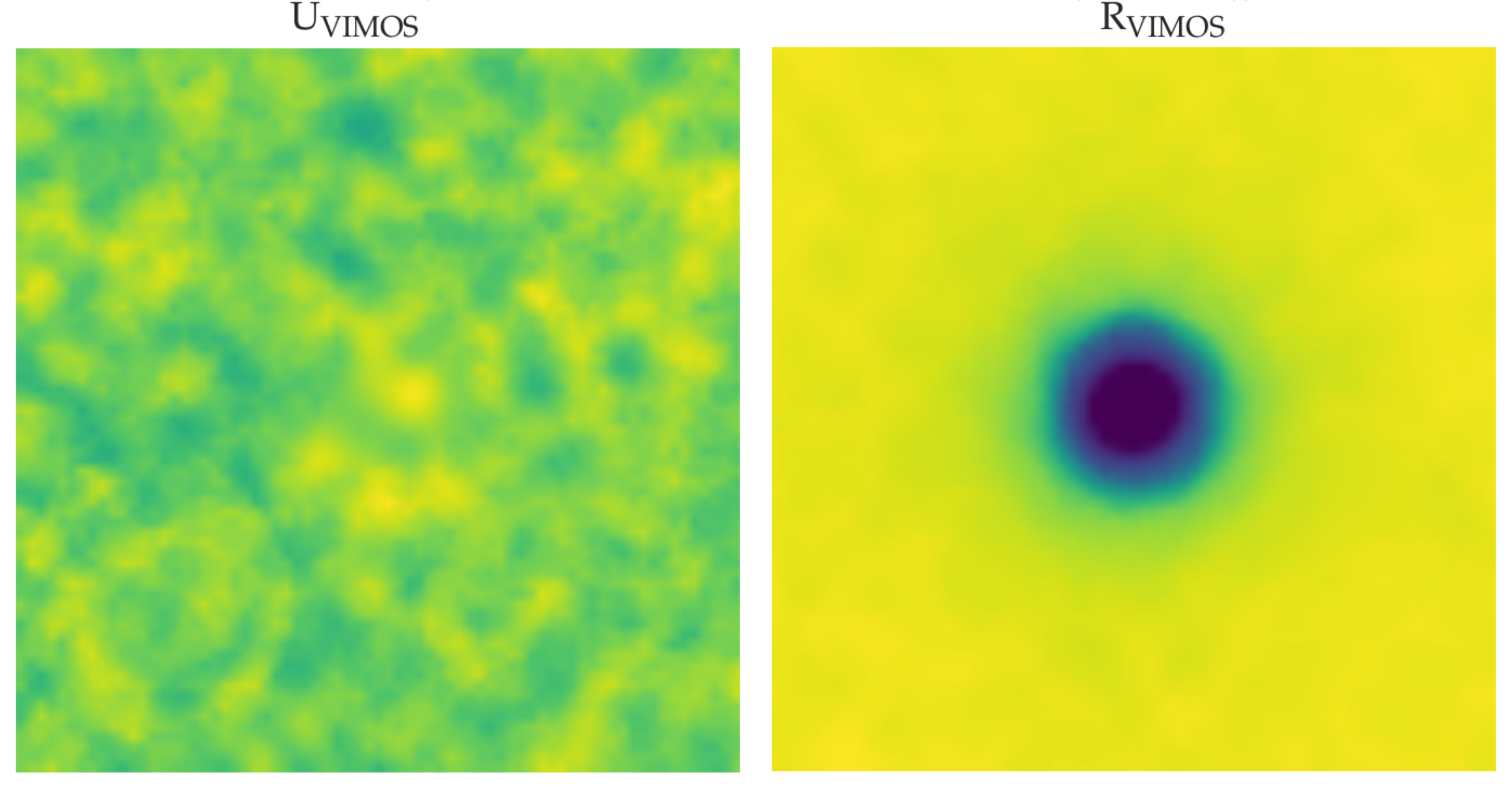}
\caption{U-band (left) and R-band (right) mean-stacks of our sample (N=73). We find no detection in the U-band stack (31.98 mag at 1$\sigma$), and place a $1 \sigma$ limit of $<6.3^{+0.7}_{-0.7}\%$ on $f_{esc}^{rel}$. The images shown are 4.3''x 4.3''.}
\label{fig:stack}
\end{figure}

\section{Results}

\subsection{The average $f_{esc}^{rel}$ at $z=3.5$}

\begin{table}
\centering
\caption{Summary of $f_{esc}^{rel}$ constraints.}
\label{summary_table}
\begin{threeparttable}
\begin{tabular}{lcc}
\hline
\multicolumn{1}{l}{Sample [N]} & \multicolumn{1}{c}{$R_{mag} (AB)$} & \multicolumn{1}{c}{$f_{esc}^{rel}$\% ($1\sigma$ upper limit)}\\
\hline
\noalign{\smallskip}
$\mathrm{Bright\tnote{\ (a)}\ \ \ \ \ \  [37] }$   & 25.03 & $6.5^{+0.7}_{-0.7}$ \\[4pt] 
$\mathrm{Faint\tnote{\ (a)}\ \ \ \ \ \  [36] }$   & 25.97 & $15.3^{+1.7}_{-1.7}\%$ \\[4pt]
\hline
All [73] & 25.38 & $6.3^{+0.7}_{-0.7}$\\[4pt]
\hline
EELGs\tnote{\ (b)}\ \ \ \ \ \  (bright) [23] & 25.07  & $8.5^{+1}_{-1}$\\[4pt]
EELGs\tnote{\ (b)}\ \ \ \ \ \   (faint) [31] & 25.96 & $16.7^{+1.8}_{-1.8}$\\[4pt]
\hline
All EELGs [54] & 25.48 & $8.2^{+0.8}_{-0.8}$\\[4pt]
\hline
\noalign{\smallskip}
\end{tabular}
\begin{tablenotes}
\item[](a) Bright: $R_{mag} < 25.38$, Faint: $R_{mag} > 25.38$. $R_{mag}=25.38$ for the Full Sample stack.
\item[](b) The EELG sample in this paper has an average estimated rest-frame EW(\OIII$\lambda5007\angstrom$) of  $\sim 400\angstrom$ as per the composite SED-estimated EWs in Table 1 of \citet{Forrest17}, who used K-band flux excesses to infer extreme \OIII\ emission.
\end{tablenotes}
\end{threeparttable}
\end{table}

\begin{figure*}
\centering
\includegraphics[width=0.95\linewidth]{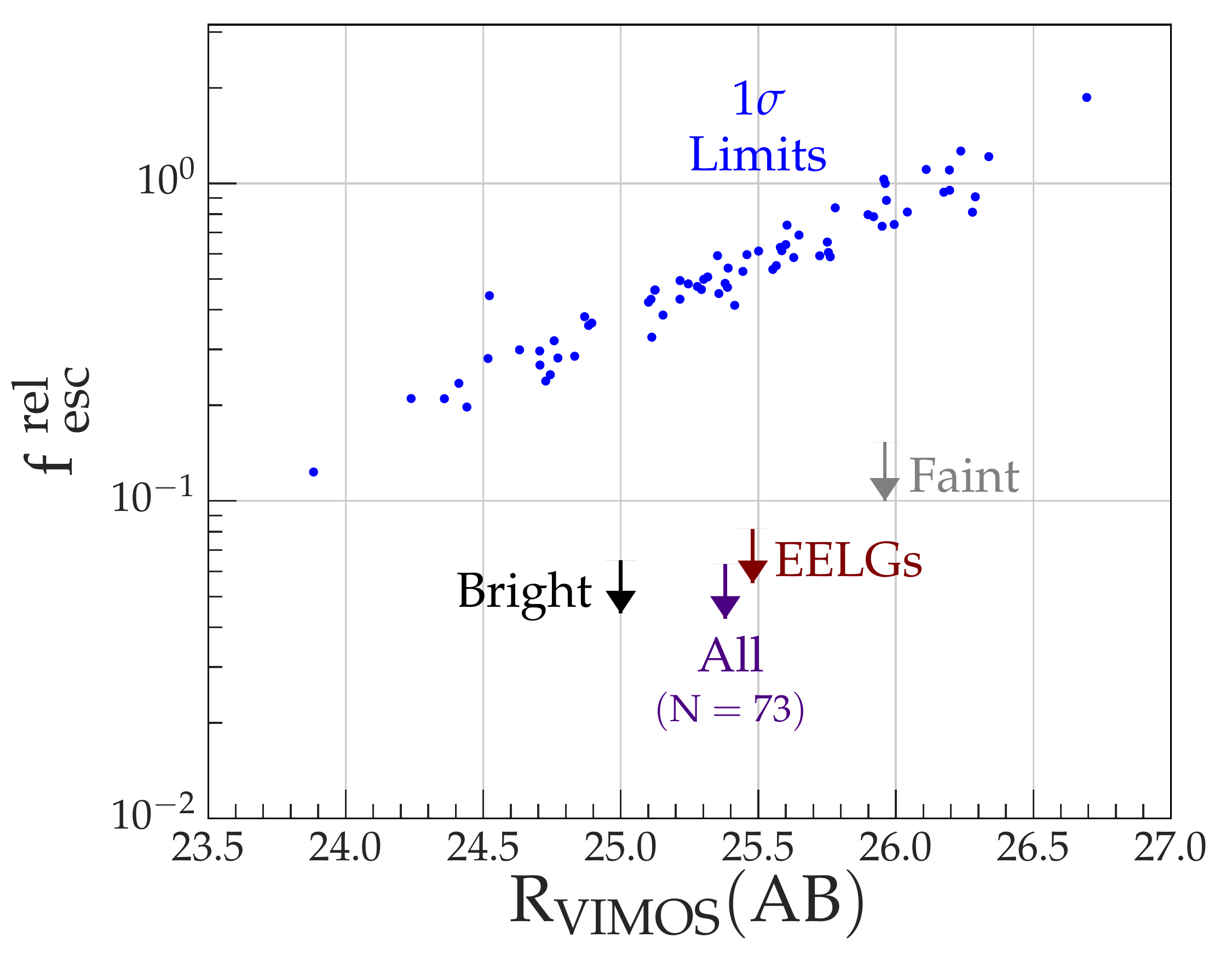}
\caption{$1 \sigma$ upper limits on $f_{esc}^{rel}$ for individual galaxies (blue) plotted against R-band magnitude. $1 \sigma$ limits on the stacked samples, ``Bright", ``Faint", ``EELGs", and ``All" defined as per Table 1, are shown in black, grey, maroon, and indigo respectively. Faint galaxies have been proposed as the drivers of reionization at $z>6$. The upward trend in our plot among the blue points should not be viewed as support for this hypothesis, but as an outcome of increasing uncertainty in the flux as the sources get fainter. Also note that the $f_{esc}^{rel}$ value for our extreme \OIII-emitter dominated stack lies well below the $10^{-1}$ mark that reionization calculations conclude is the minimum that star-forming galaxies should have, if they are the chief drivers of reionization.}
\label{fig:rmag}
\end{figure*}

\begin{figure}
\centering
\includegraphics[width=1.05\linewidth]{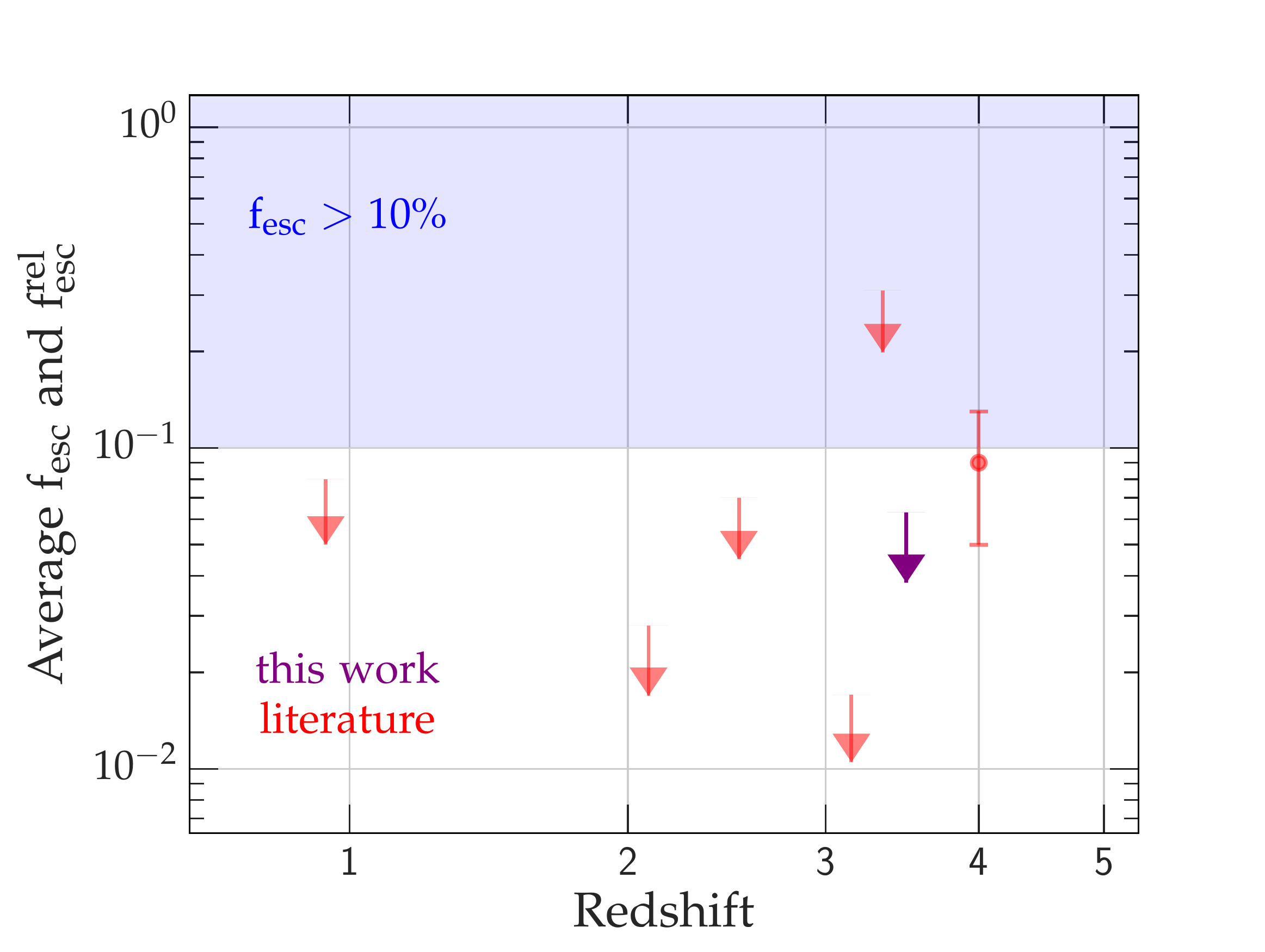}
\caption{A compilation of recent estimates of $f_{esc}^{rel}$ and $f_{esc}$ from various stacks that include generic star-forming galaxies as well as sub-samples expected to have high $f_{esc}$ \citep{Matthee16, Grazian17, Rutkowski16, Rutkowski17, Marchi16, Japelj17}. It is becoming clearer that typical star-forming galaxies are incapable of producing the $f_{esc}>10\%$ (the blue region in the plot) required by most reionization calculations in order for star-forming galaxies to drive reionization (the limit from \citet{Marchi16} at $z\sim4$ is on $f_{esc}^{rel}$, which is $\geq f_{esc}$, and \citet{Japelj17}'s extremely faint sample explains their loose bound at $z\sim3.5$). Further, even special sub-samples of star-forming galaxies expected to have high $f_{esc}$ fail to breach the $10\%$ limit-- galaxies with $\log(M/M_\odot)\sim9$ at $z\sim1$ \citep{Rutkowski16}, extreme \OIII\ emitters at $z\sim3.5$ (this work), emission-line galaxies at  $z\sim2.7$ \citep{Rutkowski17}, H$\alpha$-emitters at  $z\sim2.2$ \citep{Matthee16}). These observational estimates are in agreement with theoretical predictions that there is no appreciable evolution in the average $f_{esc}$ with redshift.}
\label{fig:z_summary}
\end{figure}

We create a U-band and R-band mean stack of our sample (shown in Figure \ref{fig:stack}). The U-band covers the LyC region at $z=3.5$ and the R-band contains non-ionizing UV radiation. While performing the stacking we take care to mask neighbouring sources and background noise using segmentation maps from the 3D-HST survey. Fluxes are extracted in 1'' apertures following \citet{Nonino09}, and an aperture correction calculated by the growth curve method \citep{Howell89} is applied to calculate the total flux.

As shown in Figure \ref{fig:stack} we find no detection in the U-band stack (which translates to a limit of 31.98 mag at $1\sigma$ for 73 sources), and a clear detection in the R band stack (25.38 mag, with S/N$>300$) and hence calculate $f_{U}/f_{R} \lesssim 1/435$. For the $1\sigma$ limit on the U-band non-detection for a single source, we measure the RMS of 10,000 sky positions in 1'' (radius) apertures and get a value (29.65 mag) consistent with \citet{Nonino09} (29.8 mag). And then for the stack $1\sigma$ limit, we divide the derived limit by $\sqrt{N}$, where $N$ is the number of sources in the stack ($N=73$ for the full stack, see Table 1 for sub-stacks).

To account for variation in the IGM in the full stack we sample $N=73$ random lines of sight (for various sub-stacks we sample lines of sight corresponding to the number of sources in each sub-stack) a million times from the 10,000 lines of sight described in Section \ref{sec:methods} to calculate $<T(IGM)_{U}> = 0.18^{+0.02}_{-0.02}$. Using these values in Equation 1, we are able to place a stringent $1 \sigma (2 \sigma)$ limit of $<6.3^{+0.7}_{-0.7}\% (<12.6^{+1.4}_{-1.4}\%)$ on $f_{esc}^{rel}$ for our sample spanning $R_{mag}=23.8-26.9$. Our result is consistent with \citet{Grazian17}, the most directly comparable study in terms of sample size and redshift selection, who place a limit of $f_{esc}^{rel}<1.70 \pm 0.14\%$ by stacking 69 galaxies at a proximal redshift range ($z=3.27-3.4$) using a shallower U-band stack (31.74 mag at $1\sigma$) at a higher IGM/U-band transmission (0.28) and with a significantly brighter sample--$R_{mag}=24.73$ compared to $R_{mag}=25.38$ in this work.

As shown in earlier work, the constraints on $f_{esc}^{rel}$ depend on the brightness of the sample. So we split the sample into two sub-stacks of ``bright" ($R_{mag}<25.38$) and ``faint" ($R_{mag}>25.38$) galaxies. We find no U-band detection in these sub-stacks as well (see Table 1 for $f_{esc}^{rel}$ constraints). In Figure \ref{fig:rmag} we show the $f_{esc}^{rel}$ values for individual galaxies in the stack. As the galaxies grow fainter, the constraint on $f_{esc}^{rel}$ grows looser since the errors on the flux are correlated with the faintness of the sources  (see Section 3 in \citet{Japelj17}).

\subsection{A first constraint on the average $f_{esc}^{rel}$ of EELGs at $z>3$}
\label{sec:first_constraint}

We create bright and faint sub-stacks of extreme \OIII\ emitters motivated by the growing speculation in the literature that \OIII/\OII\ may be used as an indirect measure of the escape fraction (see \S\ref{sec:Discussion} for more details). We use the ``Strong Emission" and ``Extreme Emission" classification from \citet{Forrest17} to select 54 sources with high \OIII\ equivalent widths from our main stack. The extreme emitters and strong emitters in \citet{Forrest17} have on average an EW(\OIII+H$\beta$) of  $\sim 803 \pm 228\angstrom$ and $\sim 230 \pm 90 \angstrom$ respectively. Our 54 selected sources have an average rest-frame EW(\OIII$\lambda5007\angstrom$) $\sim 400\angstrom$ (based on EWs derived from composite SED-fitting, see Table 1 in \citet{Forrest17}) and we refer to them as the "EELG" (Extreme Emission Line Galaxy) sample, since they have EWs comparable with EELGs reported in literature that have EW(\OIII$\lambda5007\angstrom$) ranging from $100\angstrom-1600\angstrom$ \citep{Maseda13, Maseda14, vanderWel14, Amorin15, Atek11}.

We find no detection in the U-band for the EELG sub-stacks as well, and place $1\sigma$ upper limits of $8.5^{+1.0}_{-1.0}\%$ and $16.7^{+1.8}_{-1.8}\%$ on $f_{esc}^{rel}$ for the bright and faint EELG samples respectively. Assuming an\OIII$\lambda5007$/H$\beta$ ratio of 5.1 based on \citet{Holden16}, who measure\OIII$\lambda5007$/H$\beta=5.1^{+0.5}_{-0.5}$ from a sample of star-forming galaxies at $z\sim 3.2-3.7$, and estimating EW(\OIII) and EW(\OII) using the composite SED fits from \citet{Forrest17} (see Figure 5) we roughly estimate \OIII/\OII$\sim4.3$ for our EELG sample. For a validation of EWs derived using the composite SED-fitting method see \S3.2 of \citet{Forrest17}, and note that while EW(\OIII) is exactly the same as stated in \citet{Forrest17}, EW(\OII) was estimated specifically for this work in a similar fashion. For comparison, at $z\sim3.5$, the typical \OIII/\OII$\sim2.5$ based on extrapolations from local high-$z$ analogues in SDSS \citep{Faisst16}. Given this high average ionization parameter, and the non-detection in the U-band stacks our result complicates the conclusions of recent studies like \citet{Faisst16} and \citet{Nakajima16} which point to the population of extreme \OIII\ emitters as being key to understanding reionization.

\section{Discussion}
\label{sec:Discussion}
\subsection{Evidence against a definitive LyC - \OIII/\OII\ connection}
Motivated by the incidence of extreme \OIII/\OII\ in multiple LyC leakers \citep{Vanzella16, deBarros15, Naidu17, Izotov16b}, studies like \citet{Nakajima16} and \citet{Faisst16} have looked to cast sources that show extreme \OIII\ emission and \OIII/\OII\ as the drivers of reionization. Our sample is indeed extreme for $z\sim3.5$, with an \OIII/\OII$\sim4$, which is typical of a $z=8$ galaxy as per the scaling relations for emission line EWs derived from SDSS high-$z$ analogues \citep{Faisst16, Faisst16b}.

However, we have shown here that there is no LyC signal detected in our sample, with a stringent limit of $f_{esc} \leq f_{esc}^{rel}<8.2\%$ for our EELG sample. To our knowledge, ours is the first study that has observationally constrained the LyC $f_{esc}$ for a sample of EELGs large enough to account for the stochasticity of the IGM. We acknowledge that our EELGs, while extreme, are not as extreme as some individual LyC leakers like those in \citet{Izotov16b} with EW$(\OIII)>1200\angstrom$ and \OIII/\OII$\gtrsim5$--though we do note that our sample has EW(\OIII) and \OIII/\OII\ consistent with the average galaxy in the heart of reionization at $z\sim7-8$ as per the scaling relations for emission lines in \citet{Faisst16}.

A recent comparable result which corroborates our finding of a low $f_{esc}$ even from high \OIII/\OII\ sources comes from \citet{Rutkowski17} who reported $f_{esc}<14\%$ for a small sample of 13 \OIII/\OII$>5$ galaxies at $z\sim2.3$ from a shallower $HST$/F275W stack. Other precursors to our result exist in the literature. \citet[][]{Stasinska15} used SDSS galaxies with extreme \OIII/\OII\ and a careful analysis of photoionization models to conclude that the ionization parameter on its own is an insufficient diagnostic tool for the leakage of LyC photons and must be used along with other lines like [\ion{Ar}{3}], [\ion{O}{1}] and \ion{He}{2}, and considerations of the gas covering fraction \citep[][]{Reddy16b}. There are also cases of gravitationally lensed galaxies (lensing allows looking for the LyC in faint regimes) with extreme \OIII/\OII\ and non-detections from \citet{Amorin14} (at $z=3.4$) and \citet{Vasei16} (at $z=2.4$). More recently, \citet{Izotov17} presented a detailed spectroscopic study of some of the most extreme $z<0.07$ star-forming galaxies with EW(\OIII\ )$>2000\angstrom$ and $\OIII/\OII>20$ and came to the similar conclusion that $\OIII/\OII$ by itself is not a sufficient predictor of LyC leakage. 

We conclude this section by noting that while there may be a preponderance of \OIII/\OII$\gtrsim3$  sources in the sample of confirmed LyC leakers and candidates, a high average ionization parameter for a sample (like the one presented in this paper) may not imply LyC leakage. This finding casts doubt over approaches that hope to use the population-averaged \OIII/\OII\ as an indirect measure of $f_{esc}$ during the Epoch of Reionziation. 

\subsection{Comparison with average $f_{esc}$ measurements in the literature and the missing LyC photon crisis}

Our constraints on the LyC escape fraction are in agreement with a growing string of studies that find humble escape fractions in star-forming galaxies across $z\sim1-4$ \citep[e.g.][]{Matthee16, Grazian16, Grazian17, Rutkowski16, Rutkowski17, Marchi16, Japelj17}. These constraints even include samples of highly star-forming galaxies (e.g. H$\alpha$-emitters in \citet{Matthee16}, emission-line galaxies in \citet{Rutkowski17}).

The growing consensus from simulations is that on average, there is no strong increase with redshift in the contribution of star-forming galaxies to the ionization budget \citep[e.g.,][]{Xu16, Hassan16, Anderson17}. So if the galaxies in our sample are indeed analogous to $z>6$ reionizers, their $f_{esc}^{rel}<6.3\%$ is in tension with calculations that typically require $f_{esc}\gtrsim10-15\%$ for star-forming galaxies in order for them to be the chief drivers of reionization at $z>6$ \citep[e.g.,][]{Mitra15,Mitra16,Giallongo15,madau15,Price16,Feng16, Robertson15}.

One common solution \citep[e.g.][]{Rutkowski17} to this tension is to point to the lack of low-mass ($\log(M/M_\odot)\lesssim9$), dwarf galaxies in stacking samples. This has a solid grounding in the literature since most of the simulations referenced earlier extol the role of faint, dwarf galaxies in reionization. For instance, \citet{Wise14} find an evolution of $f_{esc}$ from $5\%$ to $50\%$ between $\log(M/M_\odot)\sim8.5-7$. Our sample of EELGs has a mean $\log(M/M_\odot)\sim9$ as per the composite SED fits in \citet{Forrest17}, making it a ``low-mass" sample, but nowhere close to $\log(M/M_\odot)\sim7$. The \zfourge\  flux completeness threshold limits the presence of intrinsically faint luminosity, extremely low-mass sources in our sample, which could possibly be why we find no LyC signal in our stacks. Similarly, none of the other robust constraints on $f_{esc}^{rel}$ come from a sample comprised of dwarf, intrinsically faint galaxies--hence authors like 
\citet{Grazian17} look at gravitationally lensed sources as the next frontier for LyC studies.

Another way out of the crisis lies in revisiting some of the fundamental assumptions of reionization calculations. Accounting for the effects of binary stellar evolution (which itself is still poorly understood) for instance, allows for as much as a factor of four higher ionizing photon production efficiency \citep{Ma16, Stanway16, Choi17}. Updating assumptions about $L_{900}/L_{1500}$ and a proper treatment of dust for the UV and LyC regions may also lead to a very different picture of reionization \citep{Reddy16a, Reddy16b, Reddy17}. There is also growing evidence from simulations that the escape of ionizing photons may be a highly stochastic process with brief periods of high $f_{esc}$ \citep[][]{Wise14, Cen15, Trebitsch17}.

\section{Summary and Outlook}

In this work we placed tight constraints on the escape fraction of ionizing radiation at $z\sim3.5$ using a recently confirmed overdensity in the GOODS-South field revealed by the \zfourge\ survey that boasts an abundance of extreme \OIII\ emitters. Galaxies with extreme \OIII\ emission and \OIII/\OII$\gtrsim3$ have been marked as prime candidates for driving reionization. Here we place a tight limit on $f_{esc}^{rel}$ for such sources using a large sample with high \OIII/\OII$\gtrsim4$ and an extreme average rest-frame \OIII\ equivalent width ($\sim400\angstrom$). Our $N=73$ galaxies lie in a narrow redshift slice at $z=3.42-3.57$ which means we have a sample large enough to compensate for the stochasticity of the IGM and are not forced to average the rapidly evolving IGM transmission across a wide redshift range to gather a statistical sample size like many previous studies.

Our main finding is that despite stacking a sample expected to have an elevated $f_{esc}$, we find no LyC signal in an extremely deep U-band stack ($\sim$32mag at $1\sigma$), which corresponds to $f_{esc}^{rel}<3.8^{+0.4}_{-0.4}$. Our result raises questions about the reliability of extreme EW(\OIII) and \OIII/\OII\ as effective tracers of the LyC $f_{esc}$. We have shown in this work that these indicators do not necessarily imply significant LyC leakage for a sample, or even for individual sources. The caveats we acknowledge are that our sample doesn't contain galaxies at extremely low masses ($\log(M/M_\odot)\sim7-8.5$) and that our EELGs are not as extreme as some individual LyC leakers.

Even other stacks of the star-forming galaxy population in the literature that are expected to have higher escape fractions (e.g. low-mass galaxies, H$\alpha$ emitters, emission-line galaxies, extreme \OIII\ emitters) fail to breach the $10-15\%$ barrier on average (though there are a handful of recently discovered individual sources with much higher $f_{esc}$). Which is to say, the open questions about the sources that drove reionization are still very much open. One clear path forward is the careful analysis of the handful of known LyC leakers in the $z\sim2-4$ universe, and in efforts to bolster this small sample. Extreme \OIII/\OII\ is a tantalizing indirect estimator because it occurs both in local LyC sources \citep[e.g.][]{Izotov16a, Izotov16b} as well as in what was the only known $z>0.3$ LyC leaker for a long time ($Ion2$ from \citet{deBarros15} and \citet{Vanzella16}). However, as the sparse sample of LyC leakers grows, better estimators (perhaps the H$\beta$/UV-Slope, or Ly$\alpha$ line profiles), or combinations of indirect estimators may emerge that will maximize the potential of JWST in probing reionization directly for galaxies at $z>6$.

\section*{Acknowledgements}
\vspace{0.2cm}

We are grateful to the referee for a thorough and fair assessment of this work--their eye for detail and precise, constructive suggestions are a heartening reminder that not all is rotten in the state of peer review \citep{Smith06}. We thank Akio Inoue for providing Monte-Carlo realizations of the IGM transmission. RN thanks the members of the Kavli Institute for Astronomy and Astrophysics at Peking University and the Yale Center Beijing, particularly Prof. Linhua Jiang and Shuyan Liu, for their warm hospitality while he was working on this project. RN expresses gratitude to Prof. Shaffique Adam (Yale-NUS College) for access to High Performance Computing resources. PO acknowledges support by the Swiss National Science Foundation through the SNSF Professorship grant 157567 ``Galaxy Build-up at Cosmic Dawn". BF and KT acknowledge the support of the United States' National Science Foundation under Grant \#1410728. This research made use of Astropy, a community-developed core Python package for Astronomy \citep{astropy} and Photutils \citep{photutils}, an affiliated package of Astropy that provides tools to conduct photometry. The plots in this paper were created with Matplotlib \citep{Hunter07} and Seaborn \citep{Waskom14}. This research was possible due to the following facilities: VLT (VIMOS), HST (ACS, WFC3), Keck (MOSFIRE). Lin-Manuel Miranda (specifically, ``Non-Stop" \citep{Hamilton}) \& Yale Spring Fling provided inspiration to write this paper with celerity.

\bibliographystyle{mnras}
\bibliography{MasterBiblio.bib}


\bsp	
\label{lastpage}
\end{document}